# Bound to succeed: Transcription factor binding site prediction and its contribution to understanding virulence and environmental adaptation in bacterial plant pathogens

Surya Saha and Magdalen Lindeberg*

Department of Plant Pathology and Plant-Microbe Biology

Cornell University

Ithaca NY 14853

*corresponding author

## Abstract

Bacterial plant pathogens rely on a battalion of transcription factors to fine-tune their response to changing environmental conditions and marshal the genetic resources required for successful pathogenesis. Prediction of transcription factor binding sites represents an important tool for elucidating regulatory networks, and has been conducted in multiple genera of plant pathogenic bacteria for the purpose of better understanding mechanisms of survival and pathogenesis. The major categories of transcription factor binding sites that have been characterized are reviewed here with emphasis on in silico methods used for site identification and challenges therein, their applicability to different types of sequence datasets, and insights into mechanisms of virulence and survival that have been gained through binding site mapping. An improved strategy for establishing E value cutoffs when using existing models to screen uncharacterized genomes is also discussed.

Bacterial plant pathogens move between diverse environmental niches during the course of the disease cycle. For free-living bacteria these can include soil, epiphytic surfaces, and different tissues within the plant host, while many fastidious bacteria, though more limited in their niche flexibility, nonetheless move between plant hosts and insect vectors. Survival and successful pathogenesis require the pathogen to sense and effectively respond to the nutritional conditions and anti-microbial stresses presented by the changing niches. Environmental changes are communicated by complex sensing and signaling pathways, with the transcriptional response ultimately mediated by regulatory proteins binding to motifs within the genome sequence. Unraveling the nature and hierarchy of the protein-DNA interactions involved in specific regulatory responses represents an important foundation for understanding pathogen biology.

Developments in sequencing technology over the past decade have produced a rapidly expanding body of data from which regulatory networks can be deduced, with computational prediction of transcription factor binding sites (TFBS) playing a central role in this process. Predominant data types include complete and draft genome sequences, transcriptome sequences, and sequences of the regions bound directly by various transcription factors. From these, TFBS prediction can be used identify co-regulated genes, distinguish direct from indirect effects when evaluating expression data, and link biologically confirmed binding sites to sequence motifs identified de novo (Zambelli et al., 2012). Recent genome wide mapping of binding sites for seven sigma factors in *Sinorhizobium meliloti* highlights the power of large scale sequencing combined with TFBS prediction in advancing understanding of organismal biology (Schluter et al., 2013).

Given the investment involved in experimental confirmation of transcriptional control, optimized strategies for computational TFBS prediction represent an increasingly important tool for generating informed hypotheses (van Hijum et al., 2009; Baumbach, 2010). Statistical approaches for TFBS prediction typically involve creation of hidden Markov models (HMMs) or position weight matrices (PWMs) from sequences of previously characterized binding sites or “training sets”, thereby integrating observed variation in the binding site into the predictive model. Candidate motifs are ranked by the extent of conservation with the initial model, though establishing a cutoff that best distinguishes biologically meaningful sites from background noise can be difficult due to assumptions in the algorithm used and composition of the target genome.

Several classes of TFBSs linked to pathogenicity and survival have been characterized and used for genome-wide screening in plant pathogenic bacteria. Among those reviewed here along with a discussion of their roles in pathogen biology, are a variety of sigma factors including the extracytoplasmic (ECF) sigma factors, other activators linked to specific classes of virulence factors, and repressors.  The extensive characterization of binding site sequences conducted over the last decade in diverse genera and for a variety of regulators represents an invaluable resource to the plant pathogenic research community. To leverage these accumulated data, characterized binding site sequences have been extracted from the figures, text, and supplemental data files of the various publications in which they are described and assembled into training sets for use in building HMMs and PWMs. To aid users in applying these models to other genome sequences, the Pred_cutoff pipeline has been developed and includes a computational step designed to empirically calibrate predictions to the composition of the target genome, providing a means for comparing results generated with different models.

Some of the best examples of how binding site characterization has advanced our understanding of plant pathogenicity are found in the regulatory characterization of the type III secretion system (T3SS) and its translocated effector proteins. The T3SS impacts virulence in multiple genera of gram negative plant pathogens ranging from a critical role in plant defense suppression by *Pseudomonas syringae*, *Xanthomonas* spp., *Ralstonia solanacearum*, and *Erwinia amylovora* to more subtle impacts on bacterial multiplication and infection initiation in *Dickeya dadantii*  (Toth et al., 2003; Kay and Bonas, 2009; Block and Alfano, 2011; Deslandes and Rivas, 2012).   Type III secretion systems are grouped into families (Cornelis, 2006; McCann and Guttman, 2008) with those encoded by *P. syringae*, *E. amylovora*, *Pectobacterium*, and *Dickeya* belonging to the HrpI family and regulated by the extracytoplasmic (ECF) sigma factor, HrpL. Members of the HrpII family, present in *R. solancearum* and *Xanthomonas* spp., are regulated by a member of the AraC type family of transcriptional regulators. In both families, genome-wide predictions of the transcription factor binding sites have made transformative contributions to our understanding of the number and nature of translocated effectors and their role in pathogenicity.

<u>HrpL: Regulation of the Type III secretion system in *P. syringae* and enteric plant pathogens:</u>

Genome wide mapping of TFBSs regulatory sites associated with Type III secretion was pioneered with *P. syringae* DC3000 and has proven an invaluable tool in the comprehensive identification of T3Es throughout the species. Like other ECF sigma factors, HrpL binds to a two-block motif spanning the -10 and -35 regions (Innes et al., 1993; Xiao and Hutcheson, 1994). Although strategies for HrpL binding site identification in *P. syringae* pv. *tomato* DC3000 (*Pto* DC3000) have been previously reviewed (Lindeberg et al., 2006), it is worth noting that this motif continues to represent one of the most important tools for rapid prediction of effector repertoires. The HrpL HMM has been used to identify effector gene repertoires in *Pph* 1448a (Vencato et al., 2006), and various strains of *P. syringae* pv. *tomato* and *P. syringae* pv. *maculicola* (Almeida et al., 2009; Cai et al., 2011), with computational identification of HrpL binding sites in over 35 sequenced *P. syringae* strains contributing to discovery of nearly 60 T3E families within the species (Baltrus et al., 2011). As additional *P. syringae* draft genome sequences are generated, HrpL binding site prediction is being used to rapidly identify effector inventories, predict novel members of the HrpL regulon, and incrementally refine annotation of draft genome sequences. As an alternative to deriving models from experimentally confirmed binding sites, candidate TFBSs can be derived de novo through probabilistic sampling of regions upstream of co- regulated genes. Although the application of this approach to *Pto* DC3000 genes upregulated in the presence of HrpL did not yield a motif model appreciably different from the original HrpL HMM (Fereirra et al., 2006), this strategy has been used to successfully derive motifs from expression and transcription factor binding data for other regulators.

T3SSs in the enteric plant pathogens *D. dadantii*, *Pectobacterium* spp., and *E. amylovora* are similarly regulated by HrpL, and genome-wide motif searches for the HrpL binding site have contributed to our understanding of the distinctly different role that Type III effectors play in these pathogens. In a study of *E. amylovora* strain Ea21, Boczanczy et al used HMMs derived from training sets of confirmed *E. amylovora* and *P. syringae* HrpL binding sites to identify 28 candidate sites upstream of predicted open reading frames (Bocsanczy et al., 2012). McNally et al identified 30 candidates in *E. amylovora* strain Ea1189 using an HMM derived from a different training set (McNally et al., 2011). Both studies led to identification of regulon members outside the main Type III pathogenicity island, previously uncharacterized in *E. amylovora* and impacting the bacterial-host interaction. In Ea21, HopX1 and HopAK1 were confirmed for HrpL-dependence and HopX1 shown to elicit a defense response in *Nicotiana tabacum*. In Ea1189, EAM_2938 was shown to be a HrpL regulated membrane protein unique to the *Erwinia* genus and contributing significantly to virulence on pear. Whole genome screens for HrpL binding sites further highlight that while the *E. amylovora* T3SS is required for pathogenicity, evolutionary dynamics between host and pathogen have not led to the proliferation of effectors observed in many other phytopathogen genera.

The T3SS is not required for pathogenicity in soft-rot nectrotrophs *D. dadantii* and *Pectobacterium*, though a role has been demonstrated during the early stages of infection (Toth et al., 2003). To evaluate the inventory of HrpL-dependent genes and candidate effectors in necrotrophs, Yang et al screened the *D. dadantii* 3937 genome sequence with an HMM derived from a training set of HrpL binding sites from diverse genera. No genes with HrpL-dependent expression were found outside the main T3SS gene cluster (Yang et al., 2010), consistent with the evolutionary trend observed in *E. amylovora*. HrpL binding sites are present in the Type III secretion gene cluster of *P. carotovorum* (Rantakari et al., 2001) but systematic evaluation of the complete genome using confirmed sites as a guide has not been reported (Bell et al., 2004).

HrpX and HrpB: Regulation of the Type III secretion system in *Xanthomonas* and *Ralstonia*

The hrpII family of T3SSs encoded by *Ralstonia* and *Xanthomonas* spp. are regulated by a complex signaling cascade unrelated to HrpL-dependent regulation, although the number of Type III effectors, their role in pathogenicity, and the significance of motif searches in uncovering the effector repertoire have many parallels with *P. syringae*. The T3SS and substrates in *R. solanacearum* and *Xanthomonas* are regulated by AraC-type regulators (HrpB in *R. solanacearum* and HrpX in *Xanthomonas*) which bind to PIP or hrpII boxes, respectively. The first *Xanthomonas* genome sequences published were screened with a model based on experimentally confirmed PIP boxes (Fenselau and Bonas, 1995; Wengelnik and Bonas, 1996), and were found to encode 15-20 candidates per genome (da Silva et al., 2002; Thieme et al., 2005). Subsequent screening of the over 13 *Xanthomonas* genomes sequences now available has yielded more than 40 effector families in this genus (Furutani et al., 2009; Jiang et al., 2009; Bogdanove et al., 2011; Jalan et al., 2011; Potnis et al., 2011), confirming the utility of this approach as a means of rapid identification of regulon members (http://xanthomonas.org).

A similar strategy has proven effective in *R. solanacearum*. Cunnac et al conducted a genome-wide search of *R. solanacearum* GMI1000 using confirmed sequences of the hrpII box sequence, followed by experimental confirmation of expression. 48 genes with hrpII boxes had HrpB-dependent expression, 34 of which lacked homology to known virulence genes. Among those discovered using this approach were members of the *GALA* and AWR effector gene families, subsequently shown to play important roles in pathogenesis (Angot et al., 2006; Sole et al., 2012). The binding site model has also been used to distinguish genes under direct HrpB control in expression analyses of *hrpB*⁻ and *hrpB* overexpressing strains (Occhialini et al., 2005).

Exhaustive mutagenesis of conserved residues has revealed that the PIP/hrpII box motif is composed of two blocks of sequence a fixed distance apart, to which the regulator is believed to bind as a dimer (Cunnac et al., 2004; Tsuge et al., 2005; Koebnik et al., 2006). Given the high level of conservation, many members of the regulon can be identified using a string search for TTCG-N18-TTCG; however, analysis of natural variation, mutation of individual nucleotides, and DNA footprinting collectively support a more nuanced model of the binding site. The 3$^{rd}$ and 4$^{th}$ base of the second block can accommodate significant variation, particularly in *Xanthomonas*, and the 5$^{th}$ base in each block exhibits significant conservation, contributing appreciably to the information content of the model (Tsuge et al., 2005; Koebnik et al., 2006) (Fig 1). The PIP box is located upstream of the -10 site with the distance between being highly conserved and necessary for regulator function (Mukaihara et al.; Cunnac et al., 2004; Koebnik et al., 2006; Furutani et al., 2009), though variation in the sequence of the -10 site limits its contribution to a conserved motif for genome screening.

<u>PvdS and AcsS: Regulation of siderophore production in *P. syringae*</u>

While regulators of type III secretion have been the focus of intensive motif-based discovery of genes involved in suppressing plant defense, survival in the host environment and successful pathogenicity are dependent on complex regulatory responses involving a diverse range of additional transcription factors. Among those receiving particular attention in plant pathogenic bacteria are sigma factors and transcription factors associated with two-component and quorum sensing regulatory models. For many of these, binding site characterization has been informed by high throughput expression analyses, and review of these studies highlights the complementary roles of TFBS motif identification and expression analysis in understanding the regulatory response to environmental changes.

Extracytoplasmic sigma factors (ECF) are grouped according to their roles in the cell, with those such as HrpL involved in stress response while others respond primarily to iron starvation (Helmann, 2002). PvdS, linked to regulation of the siderophore pyoverdine, is one of five iron starvation ECF sigma factors in *P. syringae*. To assess correspondence with the PvdS binding motif previously characterized in *P. aeruginosa* (Leoni et al., 2000; Wilson and Lamont, 2000), upstream regions of PvdS regulated genes in *P. syringae* were screened for the presence of conserved sequences using Gibbs sampling and found to contain conserved sequences in the promoter region with similarity to the *P. aeruginosa* PvdS binding site. HMMs based on confirmed binding sites from the two species were compared with an HMM based on the combined set. Interestingly, the model derived from the combined set proved most effective at identification of PvdS regulated genes in other *Pseudomonas* species (Swingle et al., 2008).

Binding sites for different ECF sigma factors exhibit varying degrees of similarity in their consensus motifs. While the binding sites of HrpL and PvdS are highly dissimilar, other sigma factors bind to similar sites and exhibit significant cross-regulation (Mascher et al., 2007). A recent study describes characterization of the *P. syringae* pv *syringae* B728a ECF sigma factor AcsS which regulates genes involved in synthesis and uptake of the siderophore achromobactin as well as those involved in synthesis of Psl exopolysaccharide and mangotoxin, linked to epiphytic fitness. Although a consensus binding motif has yet to be reported, the observation that several *P. syringae* pv *syringae* B728a genes in the PvdS regulon are also subject to AcsS regulation suggests that their binding sites may be similar, allowing for coordinated up-regulation of siderophore production by the two Fe-responsive regulators (Greenwald et al., 2012).

<u>Sigma factor regulons in *Xylella*</u>

Unlike the majority of bacterial plant pathogens which encode 7-14 sigma factors involved in mediation of organismal response to a range of environmental stimuli, *X. fastidiosa* has only four (Simpson and al., 2000), providing the opportunity for comprehensive experimental exploration of its regulatory response to the environment. *X. fastidiosa* encodes two heat responsive transcriptional regulators: RpoH, also known as alternate sigma factor 32, and the ECF sigma factor RpoE, responsive to both stress and temperature. *X. fastidiosa* genes responsive to heat were evaluated in two independent studies, with Koide et al using a PWM developed from a training set of validated RpoH binding sites to distinguish upregulated genes under direct RpoH control (Koide et al., 2006). A model based on validated RpoE binding sites was likewise used to distinguish those under primary RpoE control (da Silva Neto et al., 2007). The RpoE binding site in *X. fastidiosa* was found to be more variable than that observed for the homologous sigma factor in either *E. coli* or *P. aeruginosa*. The authors speculate that with only one ECF sigma factor, *X. fastidiosa* can accommodate more variation in its binding site without risking biologically counterproductive induction by other regulators.

Characterization of temperature responsive genes in *X. fastidiosa* is relevant to pathogenicity, as those genes induced by increased temperature include candidate virulence genes encoding toxins and required for xylan degradation and type II secretion. It is suggested that heat-related gene induction may be important for the success of pathogens like *X. fastidiosa* CVC that progress more rapidly in spring and summer (Koide et al., 2006). The RpoE regulon has also been investigated in *Xanthomonas* to explore its role in survival under stress, with binding site prediction used to distinguish between those genes under direct and indirect RpoE-mediated control, particularly important given the very wide range of gene expression impacted by a RpoE mutation including elements of the type III secretion machinery (Bordes et al., 2011). Interestingly, an RpoE binding site was found upstream of RpoH reinforcing the role of the RpoH regulon in responding to diverse types of environmental stress.

The remaining sigma factors encoded by *X. fastidiosa* are the housekeeping sigma factor, RpoD, and RpoN, regulating genes associated with various phenotypes including nitrogen assimilation, particularly important to pathogens such as *X. fastidiosa* that live in low nutrient environments. Genes under direct

RpoN regulon were identified by screening those differentially expressed in the presence and absence of nitrogen, coupled with a genome-wide screen using a PWM derived from RpoN binding sites in diverse bacteria (Barrios et al., 1999). This study provides a useful illustration of the complementary role of motif identification and expression analysis. RpoN-dependent gene expression relies on enhancer binding proteins, co-expression of which is required for induction. RpoN-dependent *glnA* expression, missed in the transcript analysis owing to apparent lack of expression of the appropriate EBP under the conditions used (da Silva Neto et al., 2010), was identified in a genome-wide screen with the binding site model.

<u>RcsB: Virulence regulation in *E. amylovora*</u>

The Rcs phosphorelay is a modified 2-component regulatory system that controls production of the polysaccharide amylovoran in *E. amylovora* (Wang et al., 2011). To assess other genes and potential virulence factors under Rsc control, Wang et al evaluated gene expression in the presence and absence of various Rcs components. In parallel, an HMM based on the sites bound by phosphorylated RcsB was developed from a set of 17 previously confirmed enterobacterial RscB binding sites. Twenty-eight genes with RcsB binding sites exhibited altered expression in response to mutagenesis of various RcsB signaling components and the pattern of their association has provided a foundation for generating hypotheses about this complex regulatory network. Potential roles in virulence of the RcsB regulated genes remain uncharacterized.

<u>VirR and Fur repressors: Binding site characterization informed by ChIP-seq</u>

Repressors also play an important role in modulating bacterial response to the environment. Quorum sensing regulators such as *P. atrosepticum* VirR generally act by repressing virulence gene expression until bacteria have reached a critical population threshold, while repressors of genes involved in nutrient acquisition, such as the ferric uptake regulator Fur, limit gene expression until the nutrient in question is required. Studies characterizing VirR and Fur binding sites highlight the contribution of ChIP-seq (Johnson et al., 2007), a widely adopted experimental approach involving binding the regulatory protein to DNA and sequencing the immunoprecipitated DNA, with de novo identification of conserved motifs performed by statistical methods (Liu et al., 2001; Siddharthan et al., 2005; Bailey et al., 2009). Unlike transcriptome analysis, direct sequencing of bound DNA can distinguish TFBSs without contamination from indirect effects. Nonetheless, computational predictions with defined models remain an important tool for validation and refinement of de novo motifs as well as for screening uncharacterized sequences with models derived from ChIP-seq data. ChIP-seq was performed with VirR, a repressor of cell wall degrading enzymes in *P. atrosepticum*, to identify additional genes subject to VirR regulation. Informed by the characteristics of VirR binding sites described in other bacteria, a VirR binding motif was identified from the 20 *P. atrosepticum* sequence regions most highly enriched during ChIPseq and the resulting motif used to screen the entire genome (Monson et al., 2013). In addition to repressing siderophore production and motility, VirR was shown to activate the post-transcriptional regulator, RsmA.

Identification of genes regulated by Fur in *P. syringae*, similarly involved a combination of ChIP-seq enrichment, DNA footprinting, evaluating correspondence with consensus motifs, and expression analysis (Butcher et al., 2011). Fur was shown to directly repress not only the expected Fe uptake targets but also nearly half the TonB dependent receptors in the cell. Like VirR, Fur appears capable of gene activation as well as repression. Mapping of Fur binding sites has also been conducted in *X. fastidiosa* using a model based on sequences from diverse gamma-proteobacteria for the purpose of distinguishing the subset of Fe responsive genes that might be directly regulated by Fur. Only 20% of Fe-responsive genes in *X. fastidiosa* were found to be associated with Fur boxes, indicating the presence of a larger network of Fe-responsive regulators (Zaini et al., 2008).

A new program for optimized TFBS prediction

Use of statistical methods to predict the locations of TFBSs in sequence data sets is a critical part of the analyses described here. Although particularly useful for identification of less conserved TFBSs, even those with high conservation such as the PIP/hrpII boxes can benefit from statistically based search strategies owing to significant numbers of imperfect but functional binding sites. The ability to accurately predict locations of TFBSs becomes increasingly important as a source of hypotheses concerning regulation of virulence and survival mechanisms, particularly as more genome sequences are generated using next-generation technologies and having little or no experimental characterization. TFBS predictions conducted using HMMs or PWMs are typically assigned a score and an expect (E) value, which are used as guides for distinguishing predictions with biological relevance. An E value of 1 theoretically corresponds to the score at which a given sequence is likely to occur by chance alone and thereby provides a useful metric for gauging significance. However, compositional differences among genomes can significantly impact the frequency at which a given motif occurs, such that meaningful determination of E values requires sensitive calibration to the composition of the genome being evaluated. The lack of adequate calibration represents a significant shortcoming in the computational pipelines used for screening genomes with HMMs or PWMs. The HMMER 2.3.2 program is optimized for protein models rather than nucleic acid, and the program patser, used for screening genomes with PWMs, does not include a calibration step at all.

To aid researchers in evaluating sequence data sets for the presence of binding sites previously characterized in bacterial phytopathogens, two resources have been developed. First, training sets of binding site sequences have been assembled from the figures, text, and supplemental files of the publications discussed here and are provided in a single file in the supplemental information and as individual .aln files at http://citrusgreening.org/Pred_cutoff.html. Second, a computational pipeline for binding site prediction, Pred_cutoff, has been developed from open source software modules and specifically includes a step that empirically calibrates E values to the composition of the target sequence being evaluated. As outlined in Fig 2. Pred_cutoff input data consists of the target sequence to be screened and a multiple sequence alignment of transcription factor binding sites. if users choose to screen with an HMM, Pred_cutoff uses the programs hmmbuild and hmmsearch from the HMMER 2.3.2 suite (http://www.psc.edu/general/software/packages/hmmer/Userguide.pdf) to create a model from the input training set and screen the input genome. If a PWM based screen is chosen, Pred_cutoff uses the makematrix and patser programs for model creation and screening (Thomas-Chollier et al., 2011). Calibration of E values to the composition of the input genome is achieved through incorporation of a parallel operation that empirically determines the frequency at which a given motif will occur at random given the genome's compositional properties. This is accomplished with the seq++ package which uses an optimized Markov model to generate a collection of artificial or "mock" genomes with the same size and statistical properties as the input genome (Miele et al., 2005). The programs hmmsearch or patser are run on the artificial genomes and the score at which one motif prediction occurs per artificial genome used to empirically determine the score corresponding to an E value of 1. Data output consists of a list of locations of predicted motifs with E values empirically calibrated to the composition of the genome being evaluated. The predictions can be easily imported into downstream systems such as Artemis for curation and analysis (Carver et al., 2008). The Pred_cutoff source code is designed to be useable with limited understanding of Perl. The pipeline together with detailed documentation in the README file can be accessed at the Citrus Greening –HLB Genome Resources web site (http://citrusgreening.org/Pred_cutoff/html) and at the open source code repository site Github (https://github.com/suryasaha/Pred_cutoff).

In addition to providing an E value that reflects actual genome composition, implementation of a generic method for E value computation into the pipeline provides a means for comparing the predictive value of different training sets and of different computational approaches. In a typical application, Pred_cutoff can be used to identify training sets and computational methods that most accurately predict experimentally confirmed binding sites, yielding an optimized strategy that can be applied to related genomes, with subtle differences in genome composition accounted for in the E value determination.

While Pred_cutoff facilitates extrapolation of motif prediction from experimentally characterized to uncharacterized systems, users should be mindful that predictions will be more reliable when conducted between more closely related genomes where the likelihood of motif conservation is higher and where genome composition has not undergone dramatic shifts such as the 20-30% shifts in %GC that can occur upon genome reduction. Additionally, motifs with lower information content such as the binding sites for RpoH and RpoE will generate larger numbers of spurious predictions than those with high information content such as binding motifs for HrpB, HrpX, and HrpL. The Pred_cutoff pipeline includes the option of limiting predictions to user-defined windows upstream of gene start sites, a feature which can help to distinguish biologically relevant sites. However, given the high error rates in start site prediction (Hyatt et al., 2010), exercising this option with poorly annotated genomes may eliminate biologically important predictions.

## Conclusions

Prediction of TFBSs in sequence data sets has contributed significantly to our understanding of how plant pathogenic bacteria respond to and survive in various environments. Central to interpretation and synthesis of genome sequence data, expression data, and ChIP-seq analyses, TFBS prediction has been used to identify co-regulated gene repertoires, uncover regulatory hierarchies, and enable comparisons of regulatory networks among strains and species. As increasingly volumes of sequence data become available for experimentally uncharacterized and less tractable systems, optimized prediction methods will continue to provide a valuable foundation for hypothesis development.

Acknowledgements

We would like to thank David J. Schneider for helpful advice on refining the Pred_cutoff pipeline. This work was supported by Citrus Research and Development Foundation grant FCATP08 #123 and National Science Foundation grant IOS-1025642.

Figures

| TF | Organism | regulates | type | logo of binding motif | source |
|---|---|---|---|---|---|
| HrpL | *P. syringae* | T3SS/vir | ECF σ | 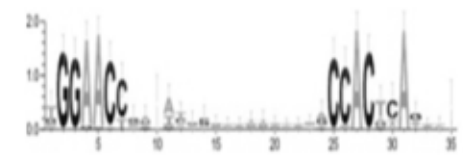 | Fouts, 2002 |
| HrpL | *E. amylovora* | T3SS/vir | ECF σ | 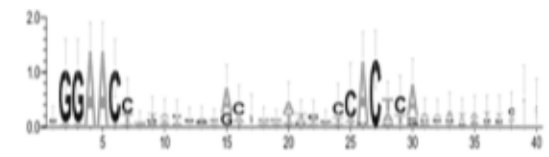 | Bocsanczy, 2012; McNally, 2011 |
| HrpL | *D. dadantii* | T3SS/vir | ECF σ | 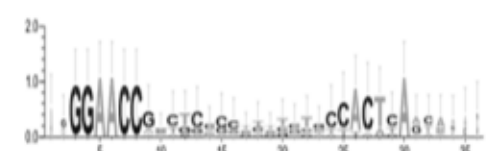 | Yang, 2010 Cooksey & Shi, 2009 |
| HrpB | *R. solanacearum* | T3SS/vir | AraC | 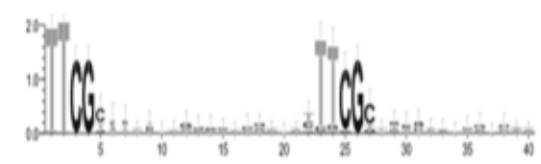 | Cunnac, 2004 Mukaihara, 2010 |
| HrpX | *Xanthomonas* spp | T3SS/vir | AraC | 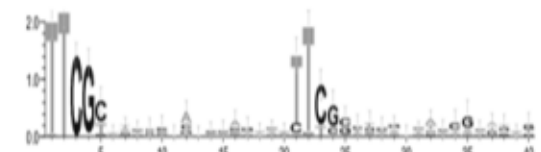 | Koebnik, 2009, Qian, 2004 Tsuge, 2005, daSilva, 2002 |
| RpoN | *X. fastidiosa* | N assimilation | σ54 | 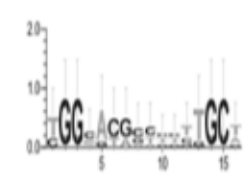 | daSilva, 2010 |
| RpoH | *X. fastidiosa* | heat | σ32 | 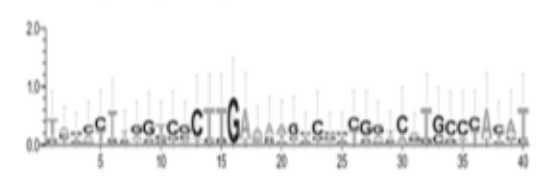 | daSilva, 2007 |
| RpoE | *X. fastidiosa* | heat/stress | ECF σ | 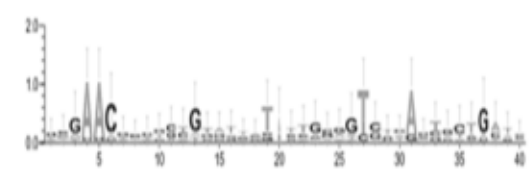 | Koide, 2006 |
| RpoE | *X. campestris* | heat/stress | ECF σ | 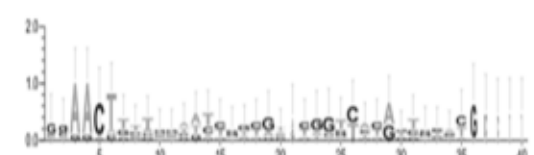 | Bordes, 2011 |
| PvdS | *P. syringae* | Fe assimilation | ECF σ | 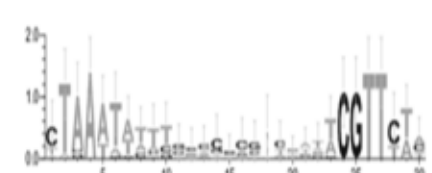 | Swingle, 2008 |
| VirR | *P. atrosepticum* | CW degrading enzymes | repressor (quorum-sens) | 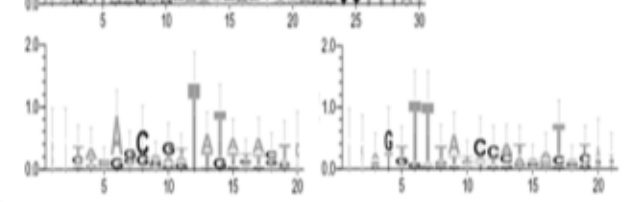 | Monson, 2012 |
| Fur | *P. syringae* | Fe assimilation | repressor | 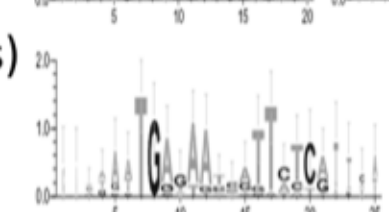 | Butcher, 2011 |
| Rcs | *E. amylovora* | amylovoran | 2-component | 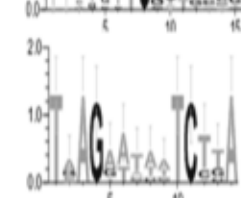 | Wang, 2012 |

FIG 1. Transcription factor binding sites characterized in plant pathogenic bacteria including the name of the transcription factor (TF), the organism in which it was characterized, its role in the cell, the transcription factor type, the logo of the binding site, and the source of sequences used in logo generation.

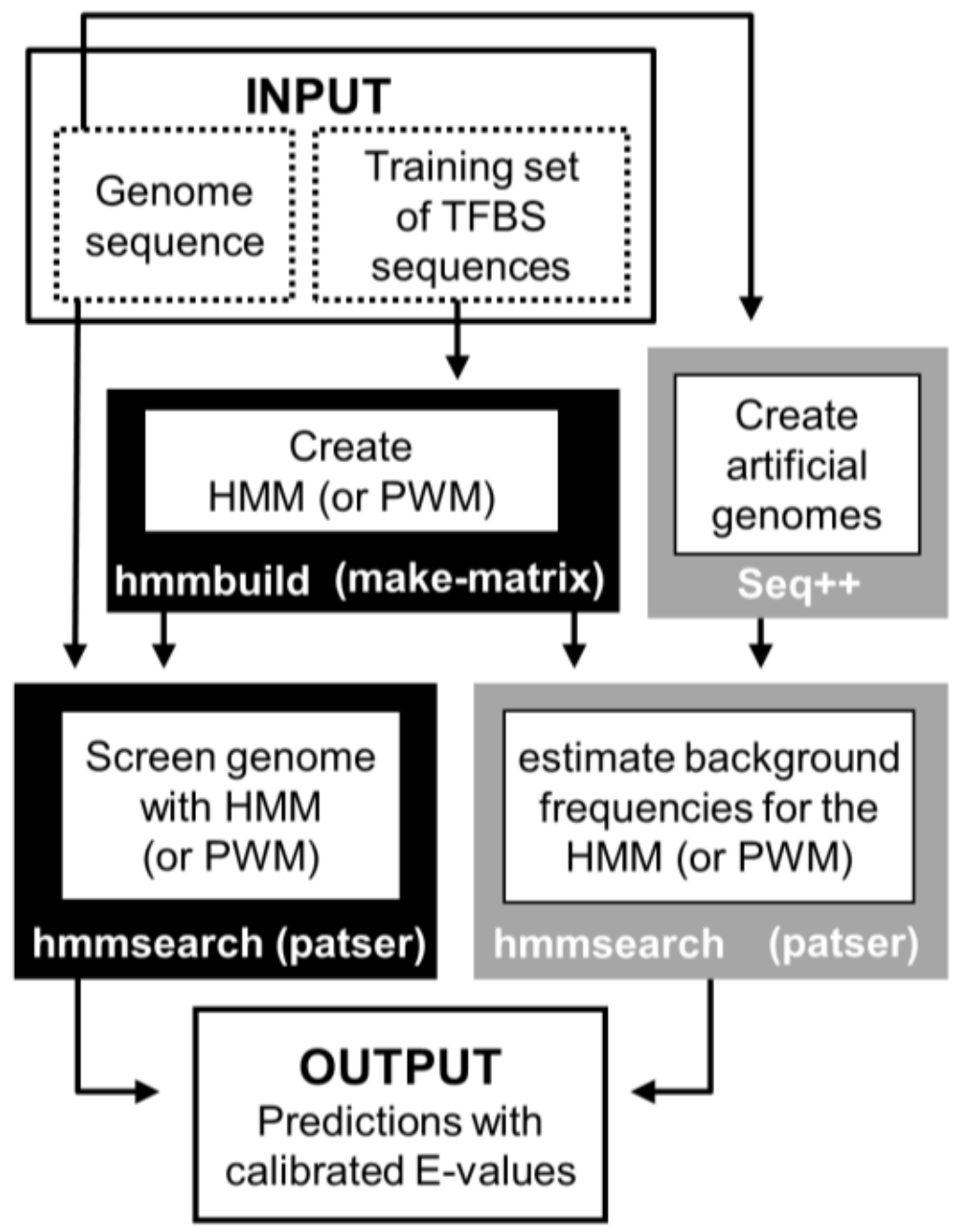

FIG 2. Diagram of the Pred_cutoff pipeline for binding site prediction with empirically calibrated E values. The genome sequence and TFBS training sets are shown in the white input box. Model creation and genome screening using either an HMM or PWM are indicated in the black boxes. Genome calibration consisting of generating artificial genomes and estimating background frequency for the model is shown in the gray boxes. Output consists of motif predictions with calibrated E values